\documentstyle[12pt]{article}
\begin{document}
\title{Interaction of Non-Abelian Monopole\\ with External Field}
\author {V.G. Kiselev\\[2mm]
 {\it {\normalsize Institute of Physics, Academy of Sciences of 
Belarus}}\\
{\it  {\normalsize Minsk 220072, Belarus}}\\[2mm]
\and Ya.M. Shnir
{\thanks{On leave of absence from the
Institute of Physics, Minsk;  Alexander von Humboldt fellow.}}\\[2mm]
{\it  {\normalsize Department of Mathematics, 
Technical University of Berlin}}\\
{\it  {\normalsize 10623 Berlin, Germany}}
\date{~}
}
\maketitle
 \begin{abstract}
The interaction of a nontrivial topological field configuration with 
the external fields is considered. The approach is based on the 
calculation of the zero modes
excitation  probability. We consider the interaction
 of the t'Hooft-Polyakov monopole
 with an external weak uniform magnetic field and the field of
another monopole. 
\end{abstract}

The t'Hooft-Polyakov non-Abelian monopole \cite{HooftPol} are 
well-known
static solutions of the nonlinear Yang-Mills-Higgs field equations.  
Though a
considerable progress has been achieved in the last two decades, 
there are
still open problems concerning the dynamical properties of monopoles.
The most known results were obtained in the
Bogomolny-Prasad-Sommerfield (BPS) limit \cite{BPS} where 
the monopole
dynamics changes drastically due to the masslessness of 
the scalar field.
A calculation of the static force between two monopoles
\cite{Manton-1} and of the light scattering by a monopole \cite{BakLee}
were based on an ansatz for the time dependence of the field which 
was just
a replacement ${\vec r} \to {\vec r} - \frac{1}{2} {\vec a}t^2$ 
for the
monopole position ${\vec r}$, that was already corresponding 
to the monopole
moving with a constant acceleration ${\vec a}$.
In Ref. \cite{Manton-1} as well as
in the following papers \cite{Gold-Wali}, \cite{LOR}, the
interaction between the monopoles was considered in the region 
outside the
monopole core where the Yang-Mills fields obey the free field
equations.
However, it resonable to expect a distortion of the core of the
t'Hooft-Polyakov
monopole and a bremstrahlung of both vector and scalar fields if the
initially static monopole configuration is accelerated by an 
external field.

In this note we describe a consistent perturbative consideration of this
idea. The only lowest-order result is presented here. It shows just the
monopole acceleration expected by the Newton law. The next-to-leading
corrections will decrease ${\vec a}$ because of the radiation.
This effect will be reported elsewhere.

Let us consider the $(3+1)$D $SU (2)$  Yang-Mills-Higgs model
specified by the Lagrangian:
\begin{equation}                                  \label{Lagrang-YM}
L = \frac{1}{4} {F_{\mu \nu}^a} {F^{\mu \nu}} ^a + \frac {1}{2} 
D_{\mu }\phi ^a
D^{\mu }\phi ^a +
\frac {\lambda}{4} (\phi ^2 - a^2)^2,
\end{equation}
where
$
F_{\mu \nu}^a = \partial _{\mu} A_{\nu}^a -  \partial _{\nu} A_{\mu} ^a  +
e \varepsilon _{abc}
 A_{\mu} ^b  A_{\nu} ^c;\quad D_{\mu }\phi ^a = \partial _{\mu} \phi ^a +
e \varepsilon _{abc} A_{\mu} ^b \phi  ^c.$

The t'Hooft-Polyakov monopole configuration reads
\cite{HooftPol}
 \begin{equation}                          \label{t'Hooft-Polyakov}
A_0 ^a = 0; \quad A_k^a =  \varepsilon _{akc} \frac{r^c}{e r^2}
\left(1 -
K(\xi)\right);\quad
\phi ^a = \frac {r^a}{e r^2} H (\xi), \quad {\rm where}~ \xi = aer.
\end{equation}
We consider first its pure electromagnetic interaction with a 
weak external
homogeneous constant magnetic field $H_k^{(ext)}$. The interaction is
described by the Lagrangian
\begin{equation}        \label{inter-mono}
L_{int} =
\frac{1}{2a}\varepsilon _{kmn} F_{km}^a \phi ^a
 H_n^{(ext)},
\end{equation}
where $H_k = (0, 0, {\cal H})$.
The field equations are:
\begin{equation}                                \label{field-eq-YM}
D^{\mu} F_{\mu \nu}^a =  e \varepsilon _{abc} \phi ^b D_{\nu }
 \phi ^c + {\it F}_{\nu} ^a;
\qquad
D^{\mu} D_{\mu} \phi ^a = \lambda (\phi ^2 - a^2) \phi ^a + {\it
F}^a.
\end{equation}
where the  last terms represent the external force acting on the
configuration. They read
\begin{equation}                   \label{external}
{\it F}_{0} ^a = 0;\quad
{\it F}_{n} ^a = - \frac{1}{a} \varepsilon _{mnc} D_m \phi ^a
H_c^{(ext)}; \quad
 {\it F} ^a =  \frac{1}{2a} \varepsilon _{mnc}  F_{mn}^a H_c^{(ext)}.
\end{equation}

The key point of our approach is to treat the excitation of the zero modes
of the monopole as the non-trivial time-dependent translation of the
configuration. The probability of this excitation can be calculated from
the field equation (\ref{field-eq-YM}). To this end we expand the fields
$A_{\mu}^a, \phi^a$:
$A_{\mu}^a = (A_{\mu}^a)_0 +  a_{\mu}^a + \dots;~
\phi ^a = (\phi ^a)_0 +  \chi^a + \dots $.
The zero
order approximation gives the classical equation
\begin{equation}                                  \label{zero-equ-YM}
D^{\mu } F_{\mu \nu}^a =  e  \varepsilon _{abc} \phi ^b D_{\nu } 
\phi ^c; \quad
D^{\mu }D_{\mu } \phi ^a = \lambda (\phi ^2 - a^2) \phi ^a
\end{equation}
with the t'Hooft-Polyakov solution (\ref{t'Hooft-Polyakov}).

Note, that in the BPS limit \cite{BPS} ($\lambda \to 0$) instead of
the field equations
(\ref{zero-equ-YM}) one has simpler first order equation
$
D_m\phi^a = \displaystyle\frac{1}{2} \varepsilon_{knm}F_{kn}^a \equiv
B_m^a,
$
and field equations (\ref{field-eq-YM}) can be written in the form
\begin{equation}
\left(D_{m} - \frac {1}{a} H_m^{(ext)}\right)F_{mn}^a =  
e  \varepsilon _{abc}
\phi ^b D_{n} \phi ^c;\quad
D_{m}\left(D_{m} - \frac {1}{a} H_m^{(ext)}\right)\phi ^a = 0,
\end{equation}
which is exactly the Manton's equations for a slowly accelerated monopole
in a weak uniform magnetic field \cite{Manton-1}.

Let us consider the corrections to the 't Hooft-Polyakov solution.
To the first order, they can be found from the equations
\begin{eqnarray}                   \label{first-equ-YM}
\left( -\frac{d^2}{dt^2} + {\cal D}^2_{(A)}\right) a_n^a & \equiv &
D^{\mu } D_{\mu } a_n^a  - e^2 [(\phi^a )^2 \delta _{ab} -
\phi ^a \phi ^b ] a_n^b + e \varepsilon_{abc} a_m^b F_{mn}^c\\
&=& 2 e \varepsilon_{abc} \chi^b D_n \phi ^c + {{\it F}_{n}^a}^{(l)};
\nonumber\\
\left( -\frac{d^2}{dt^2} + {\cal D}^2_{(\phi)}\right) \chi ^a
&\equiv & D^{\mu } D_{\mu } \chi ^a - e^2
[(\phi^a )^2 \delta _{ab} -
\phi ^a \phi ^b ]\chi ^b - \lambda [2 \phi^a \phi ^b + ((\phi^a )^2
- a^2) \delta _{ab}]
\chi ^b
\nonumber\\
 &=& - 2 e \varepsilon_{abc} a_n^b D_n \phi ^c + {{\it F}^a}^{(l)},
\nonumber
\end{eqnarray}
where the superscript $(l)$ corresponds to direction of the external field
and the background gauge
$D_{\mu} a_{\mu} ^a + e \varepsilon _{abc} \phi ^b \chi ^c = 0$
is used. In the matrix notations these equations can be rewritten as:
\begin{equation}                           \label{first-equ-mon}
\left(-\frac {d^2}{d t ^2}   + {\cal D}^2 \right)f^a   =
{{\cal F}^a}^{(l)},
\end{equation}
where
$f^a ({\bf r}, t) = \left(\begin{array}{c}
a_n^a({\bf r}, t)\\
\chi ^a({\bf r}, t)\end{array} \right), \quad
{{\cal F}^a}^{(l)} = \left(\begin{array}{c}
{{\it F}_{n}^a}^{(l)}\\
{{\it F}^a}^{(l)}\end{array} \right),
$
and ${\cal D}^2$ is the matrix obtained after two functional 
differentiations
of the action with respect to the fields $A_{\mu}^a, \phi^a$
$$
{\cal D}^2 f^a = \left(
\begin{array}{cc}
{\cal D}^2_{(A)} a_n^a   & -2 e \varepsilon_{abc} \chi^b D_n \phi ^c 
\\[3pt]
2 e \varepsilon_{abc} a_n^b D_n \phi ^c  & {\cal D}^2_{(\phi)}\chi ^a
\end{array} \right).
$$

We seek for the solution of eq. (\ref{first-equ-mon}) in the form of
an expantion
$f^a ({\bf r}, t)=\sum\limits_{i=0}^{\infty}C_i(t)\zeta^a({\bf r})_i$
on the complete set of eigenfunctions $\zeta^a({\bf r})_i$ of 
the operator
${\cal D}^2$. This eigenfunctions consists of the vector- and the scalar
components:
$\zeta^a({\bf r})_i = \left(\begin{array}{c}
\eta _n^a({\bf r})\\
\eta ^a({\bf r})\end{array} \right)$
describing the fluctuations of the corresponding fields on the monopole
backgroung \cite{Rossi}.

Substitution of the expansion $ f^a ({\bf r}, t) $
into eq.(\ref{first-equ-YM}) results in
the system of the equations for the coefficients $C_i(t)$:
\begin{eqnarray}                            \label{expan-YM-modes}
\sum _{i = 0}^{\infty} \left({\ddot C}_i + \Omega_i^2 C_i \right)
{\eta_n^a({\bf r})}_i - 2 e \varepsilon_{abc} \chi^b D_n \phi ^c
=   {{\it F}_n^a({\bf r})}^{(l)};
\nonumber\\
\sum _{i = 0}^{\infty} \left({\ddot C}_i + \omega_i^2 C_i \right)
{\eta ^a ({\bf r})}_i + 2 e \varepsilon_{abc} a_n^b D_n \phi ^c
=  {{\it F}^a ({\bf r})}^{(l)},
\end{eqnarray}

Let us consider the correction to the
monopole solution, connected with the  excitation of
the zero modes  ${\zeta^a_0}^{(k)} = \left(\begin{array}{c}
{{\eta_n^a({\bf r})}}_0^{(k)}\\
{{\eta^a({\bf r})}}_0^{(k)}\end{array} \right)$ where \cite{Rossi}

\begin{eqnarray}                   \label{modeYM-zero}
{\eta_n^a({\bf r})}_0^{(k)} = F_{kn}^a =
\partial _k  A_n^a - D_n A_k^a; \quad
{\eta ^a({\bf r})}_0^{(k)} = D_k \phi^a = \partial _k \phi ^a - e
\varepsilon_{abc} \phi^b A_k^c.
\end{eqnarray}
(here index $k$ corresponds to the translation  in the direction
${\hat r}_k$).
Note that they coincide with the pure
translation quasi-zero modes of
the vector and scalar fields ~
${{\widetilde \eta_n^a} ({\bf r})}^{(k)} = \partial _k  A_n^a; ~
{{\widetilde \eta^a}({\bf r}) }^{(k)} = \partial _k \phi ^a $ 
 up to a gauge
transformation with a special choice of the parameter that is just the
gauge potential $A_k^a$ itself.

The projection of eq.(\ref{expan-YM-modes}) onto
the zero modes gives the equation:
\begin{equation}                              \label{modeYM}
{\ddot C}_0 (N_v^2 + N_s^2) = \int d^3 x {{\it F}_n^a({\bf r})}^{(l)}
{{\eta _n^a
({\bf r})}_0}^{(k)} + \int d^3 x  {{\it F}^a({\bf r})}^{(l)}
{\eta ^a({\bf r})}_0^{(k)},
\end{equation}
(note that the non-diagonal terms cancel),
where the normalization factors of the zero modes are
\begin{eqnarray}                                    \label{orto-zero}
N_v^2 = \int d^3 x \left[{\eta_n^a({\bf r})}_0^{(k)} \right]^2
 = \int d^3 x \left({ F_{kn}^a }\right)^2;
\quad
N_s^2 = \int d^3 x \left[{\eta ^a({\bf r})}_0^{(k)} \right]^2
= \int   d^3 x \left({D_k \phi^a }\right)^2
\end{eqnarray}
There is a very simple relation between the
monopole zero modes
normalization factors and the mass of the monopole $M$:
\begin{eqnarray}
N_v^2 + N_s^2 &=& \int d^3 x \left\{
\left({{\eta_n^a}_0^{(k)}}\right)^2 +
 \left({{\eta^a}_0^{(k)}}\right)^2 \right\} = \int d^3 x \left\{ \left(
F_{kn}^a \right)^2 +  \left(D_k \phi^a \right)^2  \right\}
\nonumber\\
 &=&\frac {1}{2} \int d^3 x  \left\{ \left(
F_{kn}^a \right)^2 +  \left(D_k \phi^a \right)^2  \right\} +
\frac {1}{2} \int d^3 x ~V[\phi] = M,
\end{eqnarray}
The integrals in the r.h.s. of
eq.(\ref{modeYM}) are calculable. We find
\begin{equation}
\int d^3 x {{\it F}_n^a({\bf r})}^{(l)}
{{\eta _n^a({\bf r})}_0}^{(k)} =  \frac {2 H_k^{(ext)}}{3a}\int d^3 x
D_m \phi^a B_m^a = \frac {2}{3} g H_k^{(ext)},
\end{equation}
where we take into account the definition of the monopole 
charge $g = \int  d^3 x D_m \phi^a B_m^a = 4 \pi/e$. On the simular way
\begin{equation}
\int d^3 x
 {F^a({\bf r})}^{(l)}
{\eta ^a({\bf r})}_0^{(k)} = \frac {1}{a} \int d^3 x B_l^a H_l^{(ext)}
D_k \phi^a = \frac {1}{3} g H_k^{(ext)}.
\end{equation}
We would like to stress that only zero modes along the external field
direction $H_k$ are excited. Thus, the  equation for monopole
zero modes evolution (\ref{modeYM}) can be written as
${\ddot C}_0 (N_v^2 + N_s^2) =  g{\cal H}$,
and we have just ${\ddot C}_0 =   g {\cal H}/M$.
Hence the monopole configuration under external force is moving along
the external field direction
with a constant acceleration $a = g{\cal H}/{M}$ that
correspond to the classical Newton formula with the Lorentz force
on the monopole $F = g{\cal H} = Ma$.
The radiative correction
to this relation is given by the next order of the perturbation theory.
Note that the excitation of the monopole
zero modes (\ref{modeYM-zero}) leads not only to the translation 
of the
solution but also to its time dependent gauge transformation
$ {\cal U} = \exp\left\{-\frac{a_k t^2}{2e}A_k^a
T^a\right\}$.

Instead of eq.(\ref{inter-mono}), we could use the gauge invariant
Lagrangian of the electromagnetic interaction
\begin{equation}                             \label{inter-mono-2}
L_{int} = B_k H_k^{(ext)} = \frac{1}{2a}\varepsilon _{kmn}
\left( F_{km}^a \phi ^a
- \frac {1}{ea^2}\varepsilon _{abc}\phi ^a  D_k \phi ^b   D_m \phi ^c
\right) H_n^{(ext)}.
\end{equation}
However, the additional term in eq.(\ref{inter-mono-2}) has no effect.  
Indeed,
one can see that the additional external force on the configuration 
appears in
r.h.s. of eq.(\ref{first-equ-mon}) is orthogonal  to the monopole zero 
modes
(\ref{modeYM-zero}) and the monopole interaction with the external
electromagnetic field is determined only by the first term in the
eq.(\ref{inter-mono-2}).  The physical meaning of this result is quite 
obvious,
because the second term in the gauge invariant definition of the
electromagnetic field strength  in (\ref{inter-mono-2} ) correspond to the
Dirac monopole string in the Abelian theory.

Finally, consider the interaction between two widely separated
non-Abelian monopoles with charges $g_1$ and $g_2$.
Let us suppose that the first monopole is placed at the origine
while the second one is at the large distance $R \gg r_c$, $r_c$ stands
for the core size.
To the leading order in $r_c/R$, the field of the second monopole can be
considered as an homogenious external field acting on the first one.
Thus, the electromagnetic part of the interaction
is defined by the Lagrangian (\ref{inter-mono}) as before, where
now $H_k^{(ext)} = - g_2 R_k/R^3$ and the first monopole will
move with a constant acceleration
$
a_k = g_1 H_k^{(ext)}/M = g_1 g_2 R_k/
(M R^3) = F_k/M
$.
This corresponds to the classical Coulomb force between the monopoles:
$F_k = g_1 g_2 {R_k}/{R^3}$.

As it has been
noted in \cite{Manton-1}, \cite{Gold-Wali},
\cite{LOR}, this simple picture is not valid in the BPS limit.
Indeed, in this case the scalar field is also massless and
there is the long-range forces mediated by both
the scalar and the electromagnetic fields of the monopoles.
As usually, the scalar interaction is given
by the term
$ L_{int}' = D_m {\phi^a}^{(1)} D_m {\phi^a}^{(2)} $
in addition to the pure electromagnetic one (\ref{inter-mono}).
As the Bogomolny condition gives
$D_m \phi^a = B_m^a = \phi^a B_m$ for both monopoles
and
\begin{equation}
L_{int}' = \frac{1}{a}\phi^a
D_m {\phi^a}^{(1)}
{H_k}^{(ext)} =  \frac{1}{2a}\varepsilon _{mnk} F_{mn}^a \phi ^a
H_k^{(ext)}
\end{equation}
Thus, the total Lagrangian of interaction of the couple
BPS monopoles is just double eq.(\ref{inter-mono}) in case of the
monopole-antimonopole configuration and equal to zero in case of the
monopole-monopole (or antimonopole-antimonopole) configuration.

We are grateful to the Alexander von Humboldt Foundation for the support of
this research.
One of us (Ya.S.) thanks Professor R.~Seiler for his support and 
encouragement.
Ya.S. is also very grateful to D.~Diakonov, Per Osland, V.~Petrov and 
Ja.~Polonyi
for very  helpful  discussions  and
would like to thank the University of Bergen
where a part of this work  was  done,  for  hospitality  and
support.

\end{document}